\documentclass[prl,aps,twocolumn,showpacs,amsmath,amssymb]{revtex4-1}
\usepackage{bm}
\usepackage{graphicx}
\usepackage{color}

\usepackage{dcolumn}
\usepackage{times}
\usepackage[normalem]{ulem}

\begin{document}

\title{
Observation of Majorana Quantum Critical Behavior in a\\
Resonant Level Coupled to a Dissipative Environment}

\author{H. T. Mebrahtu,$^{1}$ I. V.  Borzenets,$^{1}$ H.  Zheng,$^{1}$ Y. V. Bomze,$^{1}$ 
A. I. Smirnov,$^{2}$ S. Florens,$^{3}$ H. U. Baranger,$^{1}$ and G. Finkelstein$^{1}$}
\affiliation{$^{1}$ Department of Physics, Duke University, Durham, NC 27708}
\affiliation{$^{2}$ Department of Chemistry, North Carolina State University,
Raleigh, NC 27695}
\affiliation{$^{3}$ Institut N\'eel, CNRS and UJF, 25 avenue des Martyrs, 
BP 166, 38042 Grenoble, France}

\begin{abstract}
We investigate experimentally an exotic state of electronic matter obtained by fine-tuning to a quantum critical point (QCP), realized in a spin-polarized resonant level coupled to strongly dissipative electrodes. Several transport scaling laws near and far from equilibrium are measured, and then accounted for theoretically. Our analysis reveals a splitting of the resonant level into two quasi-independent Majorana modes, one strongly hybridized to the leads, and the other tightly bound to the quantum dot. Residual interactions involving these Majorana fermions result in the observation of a striking quasi-linear non-Fermi liquid scattering rate at the QCP. Our devices constitute a viable alternative to topological superconductors as a platform for studying strong correlation effects within Majorana physics.  
\end{abstract}


\date{\today}

\maketitle

Quantum phase transitions (QPT)---the singular change between two distinct
ground states driven by a non-thermal control parameter---are currently
attracting strong interest in widely different fields of physics, ranging from
quantum magnets and strongly correlated materials~\cite{Sachdev} to, more
recently, cold atoms~\cite{ColdAtomRev}, nanostructures~\cite{qdots2CK,C60QPT},
and particle physics \cite{strings09}. The foremost remarkable property of QPTs
is the possibility to create exotic quantum states of matter at the quantum
critical point, such as deviations from the standard Fermi liquid paradigm for
metals; these exotic zero temperature states then cause anomalous physical
properties at finite temperature~\cite{Sachdev}. Another intriguing aspect of
QPTs is their behavior under nonequilibrium conditions, such as either a sudden
quench driving the system non-adiabatically through the transition
\cite{quenchref}, or a strong perturbation provided by a large current density
as typically realized in nanoelectronic devices \cite{BiasQPT}. Despite the
ubiquity of QPTs in contemporary theoretical physics, obtaining clear
experimental signatures has been challenging. Here, we present a thorough
characterization of all facets of a QPT in a fully-tunable single-molecule
transistor~\cite{Bomze,Mebrahtu} built from a spin-polarized carbon nanotube
quantum dot connected to strongly dissipative contacts. 

We probe the regime of resonant transport through a single level using both
linear and non-linear conductance measurements. As a reference point, in the
standard case of good metallic contacts, transport occurs via a resonance of
{\it finite} width as the electronic level of the quantum dot crosses the Fermi
energy of the leads. In contrast, the presence of dissipation drives the
conductance to zero (in the limit of vanishing temperature), unless one
tunes the center of resonant level to the Fermi energy and simultaneously makes
the tunnel barriers between the dot and the two leads perfectly equal. In 
the latter case, the conductance saturates at the unitary limit, $e^2/h$ ($e$=electron
charge, $h$=Planck constant), as we recently demonstrated~\cite{Mebrahtu}. 
A quantum critical state is obtained, whose anomalous properties
are the subject of the present study.

\begin{figure}[t]
\includegraphics[width=0.95 \columnwidth]{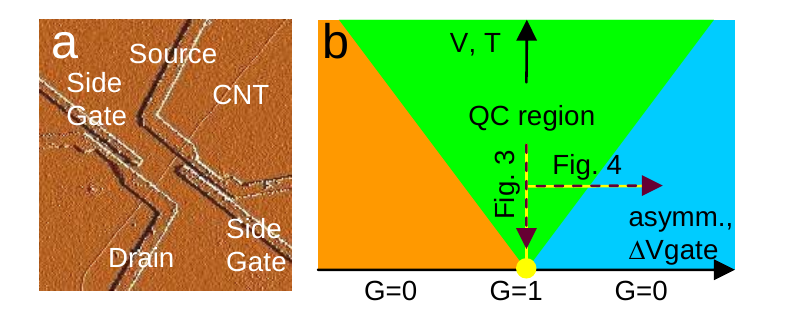}
\caption{\label{fig:Schematic}
{\bf Sample and schematic.} 
a) Atomic force microscope (AFM) image of a sample similar to the one measured.
A single carbon nanotube is connected to long and resistive source (S) and drain
(D) electrodes made of a thin Cr film. Tunnel barriers are tuned by two
lateral side gates.  
b) A diagram of the quantum phase transition in our system. The
downward flow towards the quantum critical point (marked $G\!=\!1$) is
considered in Fig.~\ref{fig:CriticalFlow}, while the runaway behavior upon gate
detuning is investigated in Fig.~\ref{fig:RunawayFlow}.
}
\end{figure}

\begin{figure*}[t] 
\includegraphics[width=1.3 \columnwidth]{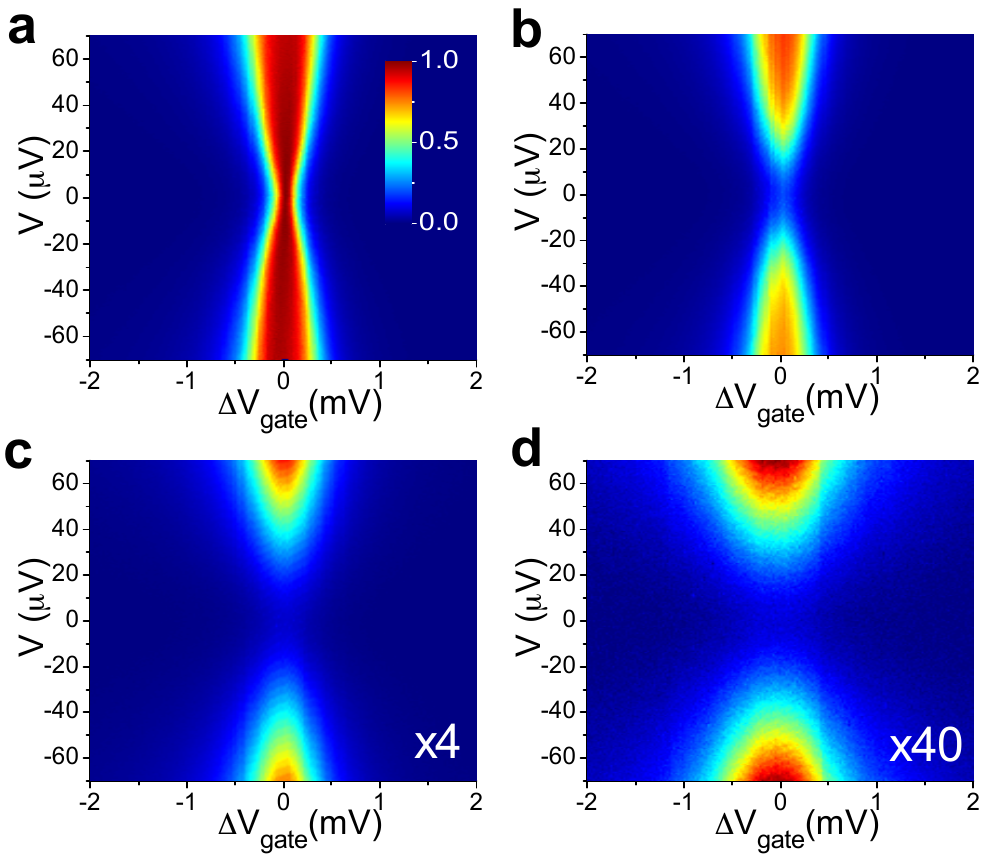}
\caption{\label{fig:overview}
{\bf Differential conductance maps close to the QCP.} 
The maps are obtained by sweeping bias $V$ and back gate offset voltage
$\Delta V_\mathrm{gate}$ (measured relative to the center of the peak) at $T
\!=\! 50$\,mK and $B \!=\! 3$\,T. Conductance in the symmetric case (a) is
followed by maps showing progressively increasing barrier asymmetry in (b-d), as
controlled by the side gate voltages. The scale of conductance maps in (c) and
(d) are multiplied by $\times 4$ and $\times 40$, respectively. The prominent
zero bias anomaly (ZBA) in the asymmetric cases completely disappears when
symmetric and on resonance, indicating a QPT. 
} 
\end{figure*}


By optimizing the dissipative environment, we find that the quantum
critical properties surprisingly behave very closely to the result for a bound
single Majorana fermion mode, causing in particular a quasi-linear scattering
rate in temperature $T$ and voltage bias $V$. Majorana fermions are presently the subject of intense scrutiny in different contexts~\cite{wilczek_majorana_2009}, ranging from particle physics~\cite{MajoranaRev1} to topological insulators~\cite{MajoranaRev2} and superconductors~\cite{
Mourik,Rokhinson,Das}.
In our case, Majorana modes emerge due to two strongly interacting leads attempting to hybridize with the
resonant level: their competition results in only partial hybridization of the level with the continuum.
The quantum critical point then corresponds to a frustrated state in which a non-hybridized Majorana mode is generated, as predicted in the related context of the two-channel Kondo model~\cite{Giamarchi,emery_kivelson}.


We study the QPT by measuring the conductance along the flow into the QCP as well as along the runaway flow to the decoupled fixed point, under both equilibrium  ($eV \!\ll\!k_BT$) and non-equilibrium ($eV \!\gtrsim\! k_BT$) conditions. This allows us to
present a comprehensive study of the scaling laws, extracting several critical
exponents which are found to be consistently determined by a single number, the
dimensionless circuit impedance, $r\equiv R e^2/h$, reaffirming our
identification of the quantum critical state.



The fabrication of the carbon nanotube quantum dot was described previously~\cite{Bomze,Mebrahtu} (see Fig.~\ref{fig:Schematic}a). The device is operated in a magnetic field of $B\!=3$ or $6$~T and at temperatures down to $T \!= 50$\,mK so that the quantum dot is effectively spin polarized.
A key feature of our samples is that the coupling of the
quantum dot to the left or right metallic lead can be varied independently using
side gate voltages. We present data measured on two resonant levels during
different measurement sessions, demonstrating compatible results. 

Differential conductance maps ($G(V)\equiv dI/dV$) near one of these resonant peaks are shown in
Fig.~\ref{fig:overview} as a function of bias voltage $V$ and gate voltage
offset $\Delta V_\textrm{gate}$ (measured from the center of the peak).
Throughout this paper, we express the differential conductance $G$ in units of
$e^2/h$. Maps for different degrees of coupling asymmetry are shown, ranging
from symmetric to very asymmetric coupling between the resonant level and the two leads. For asymmetric coupling, note the clear zero-bias anomaly
(ZBA)---a suppressed conductance at zero bias for all gate values.  The ZBA is
eliminated only for symmetric tunnel couplings and only exactly on-resonance
(Fig.~\ref{fig:overview}a). This sets the location of the quantum critical
point. These general observations \cite{Mebrahtu} are summarized in
Fig.~\ref{fig:Schematic}(b) which sketches a quantum critical diagram for our
system: at zero temperature, the system is in a conducting state with
conductance $e^2/h$ at the QCP and an insulating state
otherwise. A generic feature of QPTs~\cite{Sachdev}, which we now investigate in
our device, is the existence of quantum critical correlations close to the QCP.



\begin{figure*}[t]
\includegraphics[width=1.20 \columnwidth]{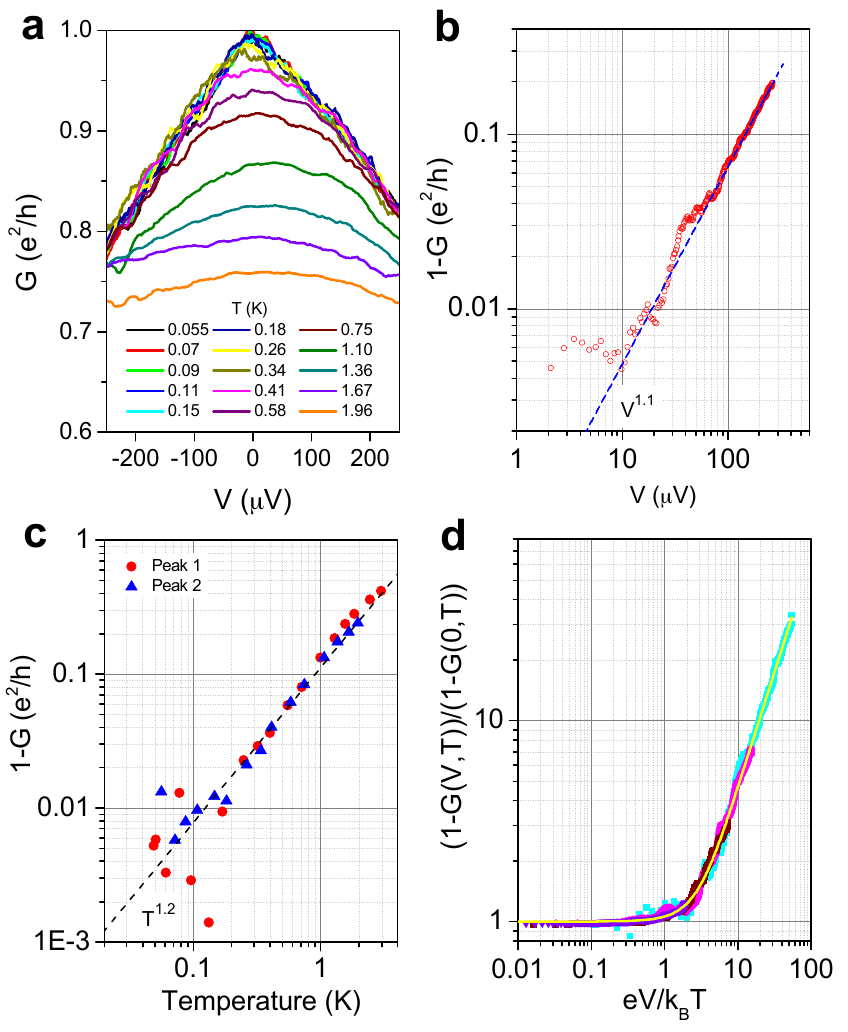}
\caption{\label{fig:CriticalFlow}
{\bf Critical flow.}
a) Nonlinear conductance $G(V,T)$ at the QCP ($\Delta
V_\textrm{gate}=0$ and symmetric barriers) measured at increasing temperatures
top to bottom: $T=$ 0.055, 0.07, 0.09, 0.11, 0.15, 0.18, 0.26, 0.34, 0.41, 0.58,
0.75, 1.10, 1.36, 1.67 and 1.96 K. The quasi-linear behavior at low $T$ signals
the presence of a Majorana state. 
b) $1-G$ measured \emph{vs.} bias voltage $V$
at $T=55$\,mK (symbols). The dashed line is a best fit to a power law $1 - G
\propto V^{\alpha}$, excluding the range $eV < k_BT$ where the conductance
saturates. 
c) Deviation of zero-bias conductance from the unitary limit $1-G$ at the
QCP as a function of temperature. Different symbols
correspond to two different Coulomb blockade peaks measured on the same sample;
the broken line is a best fit to Peak 2, giving an exponent $\alpha = 1.2$. 
d) A representative set of conductance curves from panel (a) at four different
temperatures, plotted as $1-G(V,T)$ normalized by $1-G(0,T)$. 
A line fit (see text) is superimposed on the data (symbols). The collapse of the
data onto a single curve as a function of $eV/k_B T$ demonstrates the
universality produced by proximity to the QCP. 
}
\end{figure*}

We first analyze the transport data taken with symmetric barriers and gate voltage tuned on-resonance, which corresponds to the vertical cross section of Fig.~\ref{fig:overview}a at $\Delta V_\textrm{gate}=0$. In this case, perfect transmission survives down to the lowest temperature and zero bias voltage. Fig.~\ref{fig:CriticalFlow}a shows a series of conductance \emph{vs.} bias plots measured by stepping up the temperature.
Strikingly, the conductance at the lowest temperatures shows a very unusual dependence on bias---a quasi-linear cusp. We replot $1 - G$ at the base temperature on a log-log scale in Fig.~\ref{fig:CriticalFlow}b. Clearly, the deviation of the zero-bias conductance from the unitary limit is quasi-linear: a fit to $1 -
G \propto V^{\alpha}$ yields $\alpha \approx 1.1$. Essentially the same exponent
$\alpha \approx 1.2$ is extracted from the $T$ dependence of $1-G$
(Fig.~\ref{fig:CriticalFlow}c). In the absence of dissipation in the leads, $1 -
G$ usually scales with the conventional Fermi-liquid $T^2$ or $V^2$ dependence.
Observation of a very different, intrinsically non-Fermi liquid~\cite{Giamarchi}
exponent $\alpha$ underlines the dramatic role played by the dissipative
environment.

To account for this behavior, we draw on the analogy between tunneling with
dissipation (our situation) and tunneling in a Luttinger liquid -- an
interacting one-dimensional electronic system \cite{Giamarchi}, which leads
naturally to violation of the Fermi liquid paradigm. This analogy was formally
established for tunneling through a single barrier in Ref.~\cite{SafiSaleur}; we
previously extended it to the case of spinless resonant tunneling
\cite{Mebrahtu} and will further use it here (see the Supplementary
Information). (We stress that our experiment does not probe Luttinger liquid
physics in the nanotube: at this $T$, the length of an ideal, clean nanotube
would have to be as large as $100\,\mu$m to suppress finite-size quantization.)
In the bosonization language, commonly used to describe Luttinger liquids
\cite{Giamarchi}, the approach to the QCP is controlled by
the leading irrelevant operator, $\cos (2\sqrt{\pi}\theta) \partial
\theta$~\cite{Eggert}, which corresponds to reflection of electron waves from
the charge accumulated at the resonant level. Based on the scaling dimension of
this operator, and transcribing the Luttinger interaction parameter $g$ to the
dimensionless dissipation strength $r$ according to $g=1/(r+1)$
\cite{SafiSaleur,Mebrahtu}, we expect the approach to the unitary limit to
follow $1 - G \propto = T^{2/(r+1)}$.  For this sample, the strength of the
dissipation is $r \approx 0.75$ (or $g\approx 0.57$), as measured from the zero
bias anomalies far 
off resonance~\cite{Mebrahtu}. The predicted
quantum critical exponent, $2/(r+1) \approx 1.14$, is thus very close to the
value of $\alpha$ found experimentally (Fig.~\ref{fig:CriticalFlow}b,c),
demonstrating the ability of the Luttinger liquid framework to explain our
data.


The analogy between tunneling with dissipation and tunneling in a Luttinger
liquid can provide further insight into the microscopic nature of the quantum
critical state. In the Luttinger liquid case, it is known that for $g=1/2$
the resonant level can be mapped \cite{KG} onto the quantum critical 
2-channel Kondo Hamiltonian~\cite{qdots2CK,Giamarchi,emery_kivelson}. The 
latter may in turn be
understood in terms of a hybridization of the leads with only ``half''
of the spinless fermion residing on the resonant level, leaving a fractional
degeneracy associated with the remaining non-hybridized single Majorana mode
\cite{emery_kivelson}. 
Using this intuition, we directly map our problem (resonant tunneling with dissipation) for the case 
$r=1$ and on-resonance ($\Delta V_\textrm{gate}=0$) onto the following effective Hamiltonian:
\begin{eqnarray}
\label{eq:HCritical}
H_\mathrm{Majorana}&=& 
(V_S-V_D) \big[ \hat{\psi}_f^\dagger(x=0)-\hat{\psi}_f(x=0)\big] \hat{a} \\
\nonumber
&+&  i (V_S+V_D) \big[ \hat{\psi}_f^\dagger(x=0)+ \hat{\psi}_f(x=0)\big] 
\hat{b}\\
\nonumber
&-& 2 \pi i v_F \hat{\psi}_c^\dagger(x=0) \hat{\psi}_c(x=0) \hat{a} \hat{b} .
\end{eqnarray}
Here, $\hat{a}$ and $\hat{b}$ are the two Majorana modes 
that describe a single
fermion on the dot, $\hat{d}^\dagger=\hat{a}-i\hat{b}$, and
$\hat{\psi}_{c/f}(x)$ are two extended fermionic fields 
that incorporate in a
non-trivial way the electronic degrees of freedom of the leads and the
electromagnetic phase fluctuations in the circuit. (See the Supplementary Information.)


One of the Majorana modes, $\hat{b}$, is hybridized and ``dissolves'' into the continuum. For symmetric coupling $V_S=V_D$, the first term in Eq.~(\ref{eq:HCritical}) is eliminated, and the other mode, $\hat{a}$, is not hybridized, allowing for the existence of an independent Majorana mode on the dot. The density-density interaction [last term in Eq.~(\ref{eq:HCritical})] does not destroy this independence~\cite{Sengupta,Coleman}; however, this strong
interaction (of order the Fermi energy) does govern the transport
properties near the critical point. We show~\cite{Zheng} that perturbative
treatment~\cite{Sengupta,Coleman,Zitko} of this interaction yields an anomalous
linear-in-temperature dependence, $1-G \propto T^1$~(see Supplementary Information). 

This striking deviation from Fermi liquid theory follows 
from the qualitative difference in the correlations of the hybridized and independent Majorana modes: without the third term in Eq.~(\ref{eq:HCritical}), the correlations of the hybridized Majorana fermion decrease at long time, $\langle \hat{b}^\dagger(0)\hat{b}(t)\rangle\propto 1/t$, 
while those of the non-hybridized mode do not decay, 
$\langle \hat{a}^\dagger(0)\hat{a}(t)\rangle \propto 1$.
In the calculation of the conductance, the lack of decay factor $1/t$ for the independent mode $\hat{a}$ reduces the exponent of the power-law dependence of $G$ on frequency or temperature by one, reducing the Fermi liquid dependence $T^2$ to $T^1$.
We conclude that the almost linear approach to the unitary conductance observed in Fig.~\ref{fig:CriticalFlow} 
signals the presence of a Majorana-like state.
Our experiment thus demonstrates a route to using quantum criticality to tailor
exotic electronic states in nanodevices. The rest of the paper aims to
precisely check the consistency of this scenario, and to probe the
non-equilibrium properties of this system.

An important feature of our experiment is the possibility to perturb the quantum
critical state by out of equilibrium effects, namely in the presence of a finite
voltage bias. We have already seen in Fig.~\ref{fig:CriticalFlow}c that the
dependences of $1-G$ on $T$ and $V$ feature the same exponent $\alpha$.
Interestingly, the set of curves in Fig.~\ref{fig:CriticalFlow}a associated with
the low temperature, low-bias approach to the QCP can
be rescaled onto a single universal curve, by plotting $\frac{1-G(V\neq 0,
T)}{1-G(V=0, T)}$ \emph{vs.} $eV/k_B T$ (see Fig.~\ref{fig:CriticalFlow}d). The
solid line is a numerical derivative of the heuristic expression
$\Big\vert\frac{\Gamma(\alpha+1+ieV/2\pi k_B T)}{\Gamma(1+ieV/2\pi k_B
T)}\Big\vert^2$ with respect to $V$, properly normalized to 1 at $V=0$. This
scaling function describes the experimental data quite well without any 
fitting parameters, as we use the previously determined exponent $\alpha \equiv
2/(r+1)$. This form of the scaling function is inspired by the universal scaling
of $\frac{G(V\neq 0, T)} {G(V=0, T)}$ in the regime of vanishing conductance
(away from resonance), which is given by the same functional expression with
$\alpha$ replaced by $2r$ ~\cite{Mebrahtu}.  While the scaling at weak coupling
is theoretically well established \cite{Averin_Lukens}, the scaling at the QCP,
first reported here, still requires formal justification. We note that the
similar behaviors of $G$ for the weak coupling and $1-G$ for strong coupling cannot be explained by duality of the reflection and the transmission at the corresponding fixed points \cite{KF} -- indeed, the two limiting behaviors are determined by different operators.

\begin{figure*}[t]
\includegraphics[width=1.20 \columnwidth]{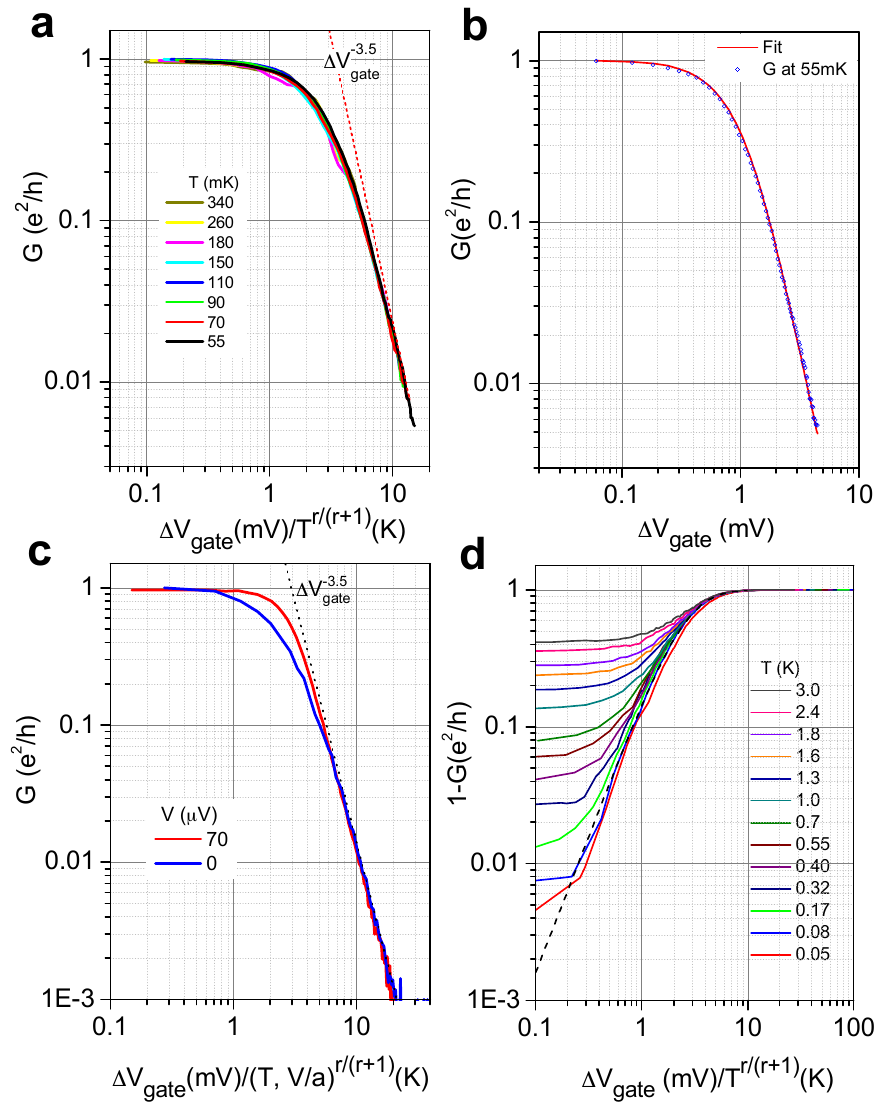}
\caption{\label{fig:RunawayFlow}
{\bf Runaway flow.}
a) Resonant peak conductance as a function of the one-parameter scaling variable 
$X=\Delta V_\textrm{gate}/T^{r/(r+1)}$. Several curves measured at eight
different temperatures between $55$ and $340$ mK scale to the same universal
curve. b) Single parameter scaling function extracted from panel (a) compared to
the universal single-barrier scaling curve, Eq.~(\ref{eq:fit}), predicted in Ref.\,\cite{AW}. A good
fit is achieved by optimizing a single fitting parameter, namely a dimensionless
prefactor accompanying $X$.  (For clarity, we plot only the data measured at the
lowest temperature.) c) Comparison of the peak shape measured at finite
$T$ under equilibrium ($eV \!\ll\! k_BT$) and non-equilibrium ($eV \!\gtrsim\!
k_BT$) conditions. The horizontal axis is $X = \Delta V_\textrm{gate}/
T^{r/(r+1)}$ in the former case and $X = \Delta V_\textrm{gate}/
(V/A)^{r/(r+1)}$ in the latter (for the value of constant $A$, see the main
text). The power laws near zero conductance and near perfect transparency are
the same, but a substantial difference between the two curves develops in the
cross-over region.  d) Behavior of $1-G$ near the peak conductance at several temperatures. Dashed line corresponds to the result of Eq.~(\ref{eq:fit}), which scales as $\sim\!X^2$ for small $X$. In each experimental curve, the change from saturation to $\sim\!X^2$ behavior marks the range associated with the QCP. Thus the quantum critical region (see Fig.~\ref{fig:Schematic}) is mapped out.
}
\end{figure*}


We now turn to the runaway behavior from the QCP toward the
zero conductance state caused by a backgate offset voltage $\Delta
V_\textrm{gate}$, as sketched in Fig.~\ref{fig:Schematic}b. We focus on the
shape of the resonant peak as a function of gate voltage and temperature or
bias. In Fig.~\ref{fig:RunawayFlow}a, the peak shape is plotted as a function of
the single parameter $X = \Delta V_\mathrm{gate}/ T^{r/(r+1)}$ \cite{KF}.
Clearly, the data sets measured at
different temperatures collapse onto a single universal curve $G(X)$.
Importantly, this scaling applies not only for high values of conductance
comparable to $e^2/h$ (as captured by the resonance width at half height
\cite{Mebrahtu}), but also to relatively large values of $X$ where $G \ll 1$.
The scaling thereby describes the full runaway flow from the strongly coupled
QCP to the weakly-coupled regime. Such single-parameter
scaling has been conjectured for resonant tunneling in the
Luttinger liquid in Ref.\,\cite{KF}, where the double-barrier structure is
treated as a single barrier with a gate-tunable transparency. The validity of
this assumption has been analyzed for weak interactions~\cite{NG,PG};
our results further corroborate the single-parameter scaling in the strongly 
interacting case ($r \sim 1$).

Concentrating first on the low conductance tail obtained for large backgate
detuning $X\gg1$, we obtain $G \propto (\Delta V_\mathrm{gate})^{-\beta}$ with
an exponent $\beta \simeq 3.4$ (Fig.~\ref{fig:RunawayFlow}c). This is very close
to the value $2(r+1) \approx 3.5$ expected from the analogy to tunneling in a
Luttinger liquid \cite{KF}. As in the case of the non-Fermi liquid exponent
$\alpha$ found in Fig.~\ref{fig:CriticalFlow} for the resonant case, this result
is markedly different from the tail of the usual Lorentzian lineshape, $G
\propto (\Delta V_\mathrm{gate})^{-2}$, expected for resonant tunneling without
dissipation. 


To further examine the assumption of single-barrier-like scaling, we compare the
universal scaling curve obtained in the experiment to the recent
non-perturbative renormalization group calculations for single-barrier
tunneling in a Luttinger liquid~\cite{AW} (Fig.~\ref{fig:RunawayFlow}b). In the
dissipation language, we use the following ansatz for the full crossover
function describing the equilibrium conductance $G$ as a function of both
detuning and temperature:
\begin{equation}
\label{eq:fit}
\frac{T}{T^\star[\Delta V_\textrm{gate}]} =
\frac{\left(G/G_0\right)^{\frac{1}{2r}}}
{\left(1-G/G_0\right)^{\frac{1+r}{2r}}}
\sqrt{\frac{1+rG/G_0}{1+r}}
\end{equation}
with $G_0=e^2/h$ and $T^\star[\Delta V_\textrm{gate}]\propto \left(\Delta
V_\textrm{gate}\right)^{\frac{1+r}{r}}$.  The excellent agreement between the
experimentally determined scaling function and the full runaway curve of Eq.~(\ref{eq:fit}) (Fig.~\ref{fig:RunawayFlow}b) vindicates the use of the single barrier expressions to describe the scaling of the double-barrier system away from the resonance.

Applying a finite bias brings additional insight by allowing us to probe the
relation between equilibrium and non-equilibrium scaling
(Fig.~\ref{fig:RunawayFlow}c).  For the Luttinger liquid system, analysis near
the decoupled fixed point~\cite{KF} predicts that the temperature scaling and
bias scaling are proportional: $b_T/b_V=\pi^{2r}[\Gamma(1+r)]^2$, where
$G(T=0,V)\simeq b_V (eV)^{2r}$ at small bias and $G(T,V=0)\simeq b_T
(k_BT)^{2r}$ at low temperature.  Thus, if we form a scaling curve using $V$ in
place of $T$, the two scaling curves should collapse onto each other in the
small conductance limit if $eV$ is divided by
$A\equiv\pi[\Gamma(1+r)]^{1/r}k_B\simeq 2.8k_B$ for $r=0.75$. The data in
Fig.~\ref{fig:RunawayFlow}c demonstrates that this holds for our system.
However, the full scaling curves are clearly quite different: though they show,
as expected~\cite{KF}, the same power laws close to both the strong and weak
coupling fixed points, the bias scaling function makes a substantially sharper
transition between the two. To the best of our knowledge, there is no firm
theoretical prediction for the full crossover curve of non-linear
conductance at arbitrary values of the dissipation parameter $r$.

Finally, quantum critical correlations are expected~\cite{Sachdev} to exist away from the critical point for temperatures larger than a characteristic gate-dependent scale (see Fig.\,\ref{fig:Schematic}). Experimentally, we can define the boundary of the quantum critical region from the behavior of $1-G$ near the peak conductance. On general grounds, a dissipation-independent exponent $1-G\propto (\Delta
V_\textrm{gate})^{2}$ is expected from the assumption of single-barrier scaling ~\cite{KF}, resulting in a universal $1-G\propto X^{2}$ dependence at small $X$. This behavior is indeed observed at low enough temperatures in Fig.~\ref{fig:RunawayFlow}d for $X \sim 1$ meV/K$^{0.43}$.
However, the approach of $1-G$ to 0 is cut off near the QCP by the leading irrelevant operator, causing $1-G$ to saturate at small $X$. (More directly, the effect of the leading irrelevant operator is seen in the power laws of Fig.~\ref{fig:CriticalFlow}.) Thereby, the saturation delineates the quantum critical region; indeed, notice that the width of the saturation region gets progressively narrower as the temperature is reduced. We may define the width of the quantum critical region in $\Delta V_\mathrm{gate}$ by comparing
the small corrections to full transparency due to the relevant operator $\propto [\Delta V_\mathrm{gate}/T^{r/(r+1)}]^2$ (Fig.~\ref{fig:RunawayFlow}) to those due to the leading irrelevant operator $\propto T^{2/(r+1)}$ (Fig.~\ref{fig:CriticalFlow}). The quantum critical region defined in this way has a simple shape: $\Delta V_\mathrm{gate} \propto T$. We convert $\Delta V_\mathrm{gate}$ to the actual energy position of the level $\Delta \epsilon_d$ by multiplying by the experimentally determined ``gate efficiency factor'' of about $0.2$. We then find $\Delta \epsilon_d / k_B T \simeq 1$, indicating that the boundary of the quantum critical region simply corresponds to the center of the conductance peak being shifted away from the Fermi energy by $\sim\!k_B T$. While appearing natural, this result is not obvious {\it a priori}, because the width of the resonant peak is much larger than $k_B T$ and scales as a non-trivial power of temperature.  

In conclusion, we presented a comprehensive study of the quantum phase
transition in a resonant level subject to finite dissipation, explicitly
demonstrating the consistency of various scaling laws based on the Luttinger
liquid analogy. This allowed us to identify the quantum critical point as
closely related to a single  Majorana bound state, with a scattering rate
quasi-linear in temperature. 
While our device, which requires fine-tuning of several gates and a carefully crafted dissipative environment, is not likely to provide a useful platform for Majorana-based quantum computing~\cite{MajoranaQC}, it certainly allows exploration  of fascinating Majorana physics in the presence of strong correlations, reaching regimes that may be harder to achieve in topological circuits. 



\begin{flushleft}{\small {\bf Supplementary Information} is linked to the online version of the paper.}\end{flushleft}

\begin{flushleft}{\small {\bf Acknowledgements.} 
We appreciate valuable discussions with  
I. Affleck, C.H. Chung, P. Coleman, K. Ingersent, K. Le Hur, and P.A. Lee.
We thank J. Liu for providing the nanotube growth facilities and W. Zhou for
helping to optimize the nanotube synthesis. H.Z., S.F., and H.U.B. thank the
Fondation Nanosiences de Grenoble for facilitating the exchange between Grenoble and Duke. 
The work in the US was supported by 
US\,DOE awards DE-SC0002765, DE-SC0005237, and DE-FG02-02ER15354.
}\end{flushleft}

\begin{flushleft}{\small {\bf Author Contributions.} 
H.T.M., I.V.B., and G.F. designed the experiment. H.T.M. fabricated the samples. H.T.M.,
I.V.B., Y.V.B., A.S. and G.F. conducted the experiment. H.T.M., H.Z., S.F.,
H.U.B., and G.F. analyzed and interpreted the data. H.Z., S.F., and H.U.B.
developed the theory.}
\end{flushleft}

\newpage

\begin{widetext}

\begin{center}
\newpage
{\large {\bf Supplementary Information for \\
``Observation of Majorana Quantum Critical Behavior in a\\
Resonant Level Coupled to a Dissipative Environment}''}\\
\bigskip
H. T. Mebrahtu,$^{1}$ I. V.  Borzenets,$^{1}$ H.  Zheng,$^{1}$
Y. V. Bomze,$^{1}$ A. I. Smirnov,$^{2}$ S. Florens,$^{3}$ H. U. Baranger,$^{1}$
and G. Finkelstein$^{1}$\\
{\it $^{1}$ Department of Physics, Duke University, Durham, NC 27708}\\
{\it $^{2}$ Department of Chemistry, North Carolina State University,
Raleigh, NC 27695}\\
{\it $^{3}$ Institut N\'eel, CNRS and UJF, 25 avenue des Martyrs, BP
166, 38042 Grenoble, France}\\
\today \\
\end{center}
\end{widetext}

\setcounter{figure}{0}
\setcounter{equation}{0}
\renewcommand{\thefigure}{S\arabic{figure}}

\section{I. Experimental details}

While several samples demonstrated qualitatively similar behavior (non-Fermi-liquid exponents close to the critical point), we chose to present a consistent set of data measured on one sample with $r$ close enough to 1. We show two single-electron peaks, labeled peaks 1 and 2 in Fig.~3c of the main text, measured during two cooldowns. Specifically, Fig.~2 and 4c,d have been measured on Peak \#1 at $B=3$ T; Figures 3 and 4a,b have been measured on Peak \#2 at $B=6$ T. The peaks happen to have similar parameters (as evidenced, \emph{e.g.} by the similar width in Fig.~4a and c), so the data measured on the two peaks can be viewed almost interchangeably. 

\begin{figure}[ht]
\includegraphics[width=1 \columnwidth]{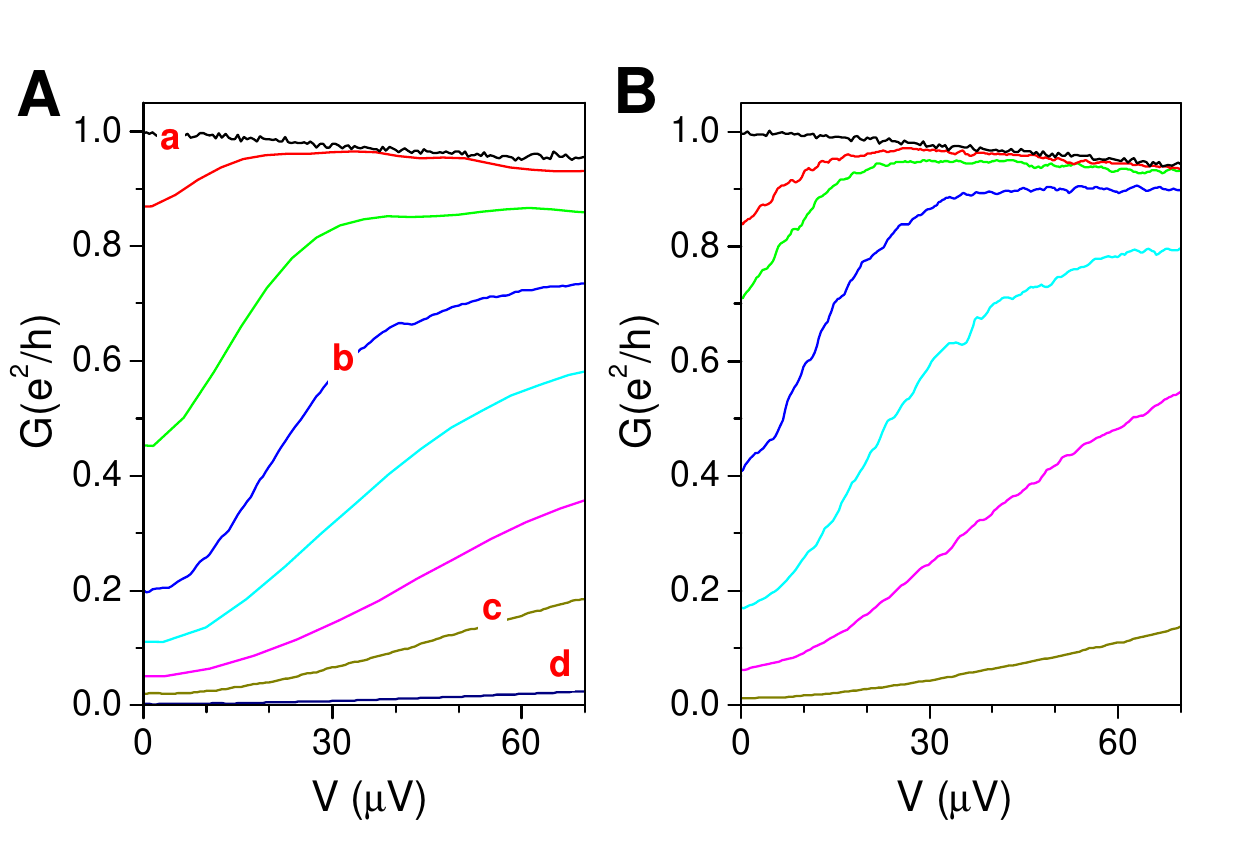}
\caption{\label{fig:cuts}
a) Nonequilibrium conductance measured at $\Delta V_\mathrm{gate}=0$ and at various
coupling asymmetries. The curves {\bf a}, {\bf b}, {\bf c}, and {\bf d}
correspond to on-resonance condition for the maps shown in Fig.2 of the main
text.
b) Conductance in the symmetrically coupled case, associated to curve {\bf a} of 
panel a), measured at increasing gate voltage $\Delta V_\mathrm{gate}$= 0, 0.7, 1.2, 
1.6, 1.7, 2.0, 5.0 mV (top to bottom). 
}
\end{figure}

Fig.~S1 represents the conductance $G$ \emph{vs.} bias voltage $V$ data corresponding to 
the vertical cross-sections of the maps in Fig.~2 of the main text. Both panels include the
curve measured at the QCP (top curve) and several curves detuned from the QPC,
either by the tunnel barrier asymmetry (panel A), or by shifting the level
off-resonance with the gate voltage (panel B). Clearly, except for the special
point of both symmetric coupling and on-resonance condition, the conductance is
suppressed at small bias. (Note that the zero-bias suppression is limited at $eV
\simeq k_B T$, leading to a saturation of the conductance at small $V$.) The
similar trends shown in both panel demonstrate the equivalent role of the gate
detuning and tunnel barrier asymmetry in destroying the quantum critical state.


In Fig.~S2, we further analyze the data of Fig.~4d and present an empirical procedure to determine the width of the quantum critical region, alternative to the one discussed in the main text. We fit conductance $G(X)$ with the eq.~(2) of the main text, replacing $X$ with an \emph{ad hoc} expression $(X^2 +X_0^2)^{1/2}$. Thus introduced width, $X_0$, is the only fitting parameter for each curve. The extracted $X_0$ is plotted in the inset and is further fit by a power law dependence on $T$. The resulting exponent of approximately $0.7$ is reasonably close to the expected $1/(1+r)$.


\begin{figure}[ht]
\includegraphics[width=0.7 \columnwidth]{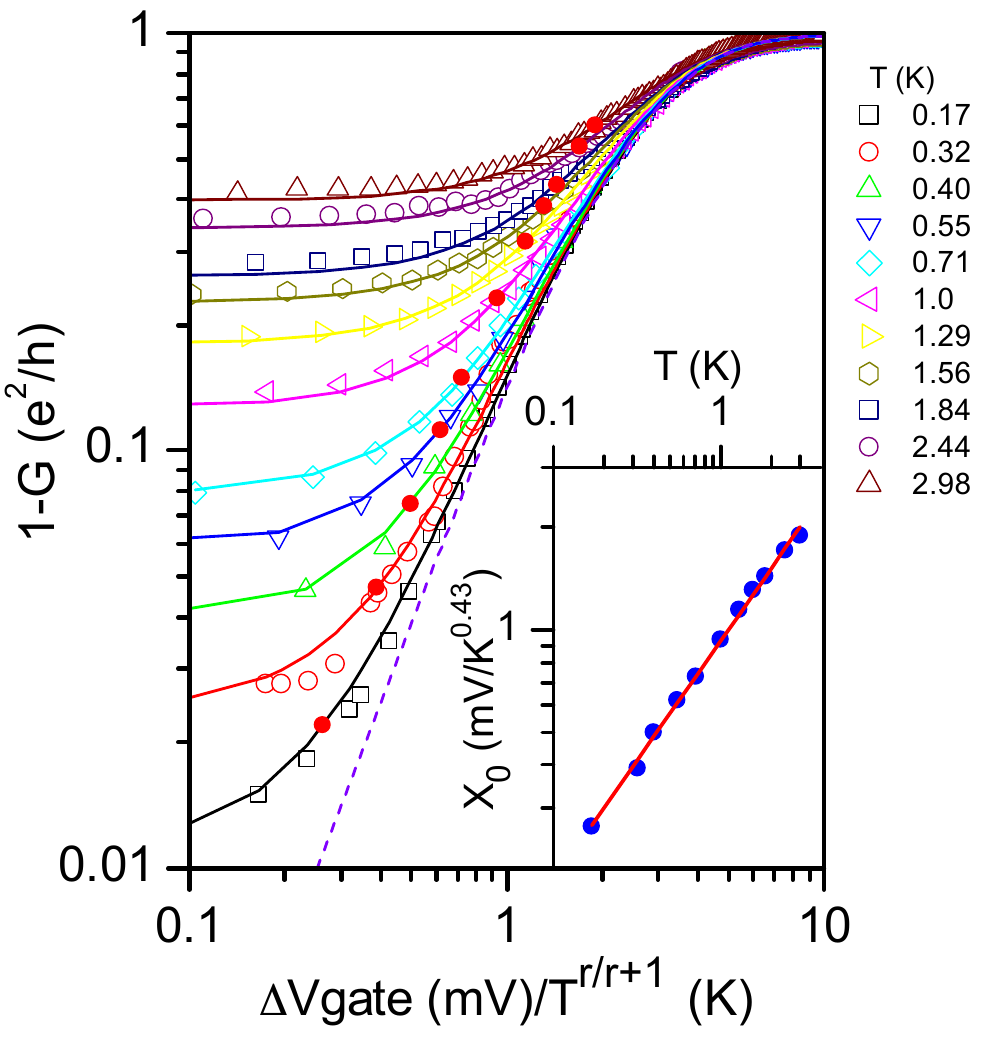}
\caption{\label{fig:CriticalReg}
a) Select sets of data from Fig.~4d fit by the expression of eq.~(2) of the main text, with $X$ empirically replaced with $(X^2 +X_0^2)^{1/2}$. Red dots indicate the fitted $X_0$ for each curve. Dashed line: eq.~(2) of the main text (without $X_0$). $X_0$ is further plotted in the inset and fit by a power law dependence on $T$. 
}
\end{figure}

\section{II. Derivation of the Hamiltonian for a dissipative spinless quantum dot}

We provide here the microscopic theory of our experiment, eventually leading 
us, in the following sections, to the
interacting Majorana resonant level Hamiltonian given in Eq.~(1) of the
main text.


Because of the large applied magnetic field, our experimental setup can be first
modeled by a spinless quantum dot, described by a fermion operator $d^\dagger$,
coupled to two conducting leads.
The key feature of our device is the presence of an Ohmic dissipative 
environment, with sizeable quantum fluctuations of the voltage in the source and drain 
leads. The total Hamiltonian thus reads: 
\begin{equation}
H=H_{\textrm{Dot}}+H_{\textrm{Leads}}+H_{\textrm{T}}+H_{\textrm{Env}},
\label{eq:H}
\end{equation}
where $H_{\textrm{Dot}}=\epsilon_{\textrm{d}}d^{\dagger}d$ is the Hamiltonian of the
dot with the energy level $\epsilon_{\textrm{d}}$ (tuned by the backgate voltage
$V_\mathrm{gate}$), and $H_{\textrm{Leads}}=\sum_{\textrm{\textrm{\ensuremath{\alpha}=S,D}}}
\sum_{k}\epsilon_{k}c_{k\alpha}^{\dagger}c_{k\alpha}$
represents the electrons in the source (S) and drain (D) leads. The most
important piece of Hamiltonian~(\ref{eq:H}) is $H_{\textrm{T}}$, which describes the 
tunneling between the dot and the leads with amplitudes $V_{S/D}$:
\begin{equation}
H_{\textrm{T}}=V_{S}\sum_{k}(c_{kS}^{\dagger}e^{-i\varphi_{S}}d+  {\rm h.c.} )
+V_{D}\sum_{k}(c_{kD}^{\dagger}e^{i\varphi_{D}}d+ {\rm h.c.} ),
\label{eq:Htunnel}
\end{equation}
where the operators $\varphi_{S/D}$ induce phase fluctuations of the
tunneling amplitude between the dot and the S/D lead. These phase operators 
are canonically conjugate to the operators $Q_{S/D}$ corresponding to charge
fluctuations on the S/D junctions. 
We follow the standard approach to treat quantum tunneling in the presence of a 
dissipative environment \cite{ingoldbook}, which is valid for electrons propagating much slower
than the electromagnetic field \cite{Nazarovbook}. 

It is useful to transform to variables related to the total charge on the dot.
To that end, we introduce \cite{ingoldbook} two new phase operators, $\varphi$
and $\psi$, related to the phases $\varphi_{S/D}$ by 
\begin{eqnarray}
\varphi_{S} & = & \kappa_{S}\varphi+\psi \qquad\nonumber \\
\varphi_{D} & = & \kappa_{D}\varphi-\psi \;,\label{eq:}
\end{eqnarray}
where $\kappa_{S/D}=C_{S/D}/(C_{S}+C_{D})$ in terms of the capacitances of the
dots to the source/drain contacts, $C_{S/D}$. The phase $\psi$ is the variable 
conjugate to the fluctuations of charge on the dot $Q_{c}=Q_{S}-Q_{D}$.
Likewise, $\varphi$ is the variable conjugate to 
$Q=(C_{S}Q_{D}+C_{D}Q_{S})/(C_{D}+C_{S})$. Assuming $C_{S}=C_{D}$, we have 
$\varphi_{S}=\varphi/2+\psi$ and $\varphi_{D}=\varphi/2-\psi$. 

The gate voltage fluctuations can be disregarded in our experiment because the
capacitance of the gate is negligible, $C_{g}\ll C_{S/D}$ (the opposite limit
of a noisy gate coupled to a resonant level was considered in
Refs.~\cite{Averin_94,LeHurQPT}). In fact, the coupling of the
fluctuations of the total charge on the dot to the environment is negligible \cite{ingoldbook}.
Thus, only the {\it relative} phase difference between the two leads remains
\cite{ingoldbook,florens07}, and the tunneling Hamiltonian becomes
\begin{equation}
H_{\textrm{T}}=V_{S}\sum_{k}(c_{kS}^{\dagger}e^{-i\frac{\varphi}{2}}d
+ {\rm h.c.})+V_{D}\sum_{k}(c_{kD}^{\dagger}e^{i\frac{\varphi}{2}}d+ {\rm h.c.} ). 
\end{equation}

The last part of Eq. (\ref{eq:H}) is the Hamiltonian of the environment,
$H_{\textrm{Env}}$ \cite{caldeira81,leggett87,ingoldbook}. The environmental
modes are represented by harmonic oscillators described by inductances and
capacitances such that their frequencies are given by
$\omega_{k}=1/\sqrt{L_{k}C_{k}}$. These oscillators are then bilinearly coupled
to the phase operator $\varphi$ through the oscillator phase:
\begin{equation}
H_{\textrm{Env}}=\frac{Q^{2}}{2C}+\sum_{k=1}^{N}\left[\frac{q_{k}^{2}}{2C_{k}}
+\left(\frac{\hbar}{e}\right)^{2}\frac{1}{2L_{k}}\left(\varphi-\varphi_{k}\right)^{2}\right]. 
\end{equation}  

\section{III. Mapping onto the Luttinger liquid tunneling problem}

In this section, we first use bosonization~\cite{GiamarchiBook} to map our 
model Eq.~(\ref{eq:H}) to that of a resonant level contacted by two Luttinger liquids. 
In carrying out this mapping, we follow closely previous work on tunneling through a 
single barrier with an environment \cite{safi04,LeHur&Li_05} and the Kondo effect in 
the presence of resistive leads \cite{florens07}.
 
The two metallic leads in our case can be reduced to two semi-infinite
one-dimensional free fermionic baths, which are non-chiral \cite{GiamarchiBook}.
By unfolding them, one can obtain two chiral fields \cite{GiamarchiBook}, which
both couple to the dot at $x=0$. We bosonize the fermionic fields in the
standard way \cite{GiamarchiBook,senechal} $c_{S/D}(x)=\frac{1}{\sqrt{2\pi a_0}}
\exp[i\phi_{S/D}(x)]$ (neglecting Klein factors for simplicity), where $\phi_{S/D}$ are 
bosonic fields introduced to describe electronic states in the leads, and 
$a_0$ is a short time cutoff. 
Defining the flavor field $\phi_{f}$ and charge field $\phi_{c}$ by
\begin{equation}
\phi_{f} \equiv \frac{\phi_{S}-\phi_{D}}{\sqrt{2}}, \qquad
\phi_{c} \equiv \frac{\phi_{S}+\phi_{D}}{\sqrt{2}},
\end{equation}
we can rewrite the Hamiltonian of the leads as
\begin{equation}
H_{\textrm{Leads}}=\frac{v_{F}}{4\pi}\int_{-\infty}^{\infty}dx
\left[\left(\partial_{x}\phi_{c}\right)^{2}+[\left(\partial_{x}\phi_{f}\right)^{2}\right]. 
\end{equation}
with $v_F$ the Fermi velocity. The tunneling Hamiltonian then becomes
\begin{eqnarray}
H_{\textrm{T}} & = & \frac{V_{S}}{\sqrt{2\pi a_0}}\,
\text{exp}\left[-i\frac{\phi_{c}(0)+\phi_{f}(0)}{\sqrt{2}}-i\frac{\varphi}{2}
\right] d+ {\rm h.c.} \nonumber \\
&  + & \frac{V_{D}}{\sqrt{2\pi a_0}}\,
\text{exp}\left[-i\frac{\phi_{c}(0)-\phi_{f}(0)}{\sqrt{2}}
+i\frac{\varphi}{2}\right] d+ {\rm h.c.}.
\end{eqnarray}
Note a key feature of $H_{\textrm{T}}$: the fields $\varphi$ and $\phi_{f}(0)$
enter in the same way in the tunnel process. Combining these two fields together
embodies  a process which is analogous to having effectively interacting leads 
as in a Luttinger liquid. 

To carry out such a recombination of the phase fields, since the tunneling only acts 
at $x=0$, it is convenient to perform a partial trace in the partition function and 
integrate out fluctuations in $\phi_{c/f}(x)$ for all $x$ away from $x=0$
\cite{KaneFisher}. For
an Ohmic environment, one can also integrate out the harmonic modes at low frequency \cite{leggett87,safi04,florens07}. Then, the 
effective action for the leads and the environment becomes
\begin{equation}
S_{\textrm{Leads+Env}}^{\textrm{eff}}=\frac{1}{\beta}\sum_{n}|\omega_{n}|
\left[|\phi_{c}(\omega_{n})|^{2}+|\phi_{f}(\omega_{n})|^{2}
+\frac{|\varphi(\omega_{n})|^{2}}{2r}\right]
\end{equation}
with $r=R e^{2}/h$ the dimensionless impedance of the leads and 
$\omega_{n}=2\pi n/\beta$ a Matsubara frequency. 
In this discrete representation, it is straightforward to combine 
the phase operator $\varphi$ and the flavor field $\phi_{f}$; in order to 
maintain canonical commutation relations while doing so, we use the transformation
\begin{eqnarray}
\phi_{f}' &\equiv&
\sqrt{g}\left(\phi_{f}+\frac{1}{\sqrt{2}}\varphi\right),\qquad \\
\varphi' &\equiv& \sqrt{g}\left(\sqrt{r}\phi_{f}
-\frac{1}{\sqrt{2r}}\varphi\right),
\end{eqnarray}
where $g\equiv 1/(1+r)\leq1$. Now, the effective action
for the leads and environment (excluding tunneling) becomes
\begin{equation}
S_{\textrm{Leads+Env}}^{\textrm{eff}}\!\!=\frac{1}{\beta}\sum_{n}|\omega_{n}|
\left(|\phi_{c}(\omega_{n})|^{2}+|\phi_{f}'(\omega_{n})|^{2}+|\varphi'(\omega_{n})|^{2}\right) ,
\label{eq:Seff}
\end{equation}
while the action for the tunneling part reads in the time-domain:
\begin{eqnarray}
\nonumber
S_{\textrm{T}}&=&\int d\tau \Bigg[\frac{V_{S}}{\sqrt{2\pi a_0}}
e^{-i\frac{1}{\sqrt{2}}\phi_{c}(\tau)}e^{-i\frac{1}{\sqrt{2g}}\phi_{f}'(\tau)}d
+{\rm c.c.}\\
&&
+\frac{V_{D}}{\sqrt{2\pi a_0}}
e^{-i\frac{1}{\sqrt{2}}\phi_{c}(\tau)}e^{i\frac{1}{\sqrt{2g}}\phi_{f}'(\tau)}d
+ {\rm c.c.}\Bigg] \;. 
\label{eq:LT}
\end{eqnarray}
Thus we see that the phase $\varphi$ has been absorbed into the new flavor field 
$\phi_{f}'$ at the expense of a modified interaction parameter $g\leq1$, while the new 
phase fluctuation $\varphi'$ decouples from the system. 
It turns out that one obtains a very similar effective action by starting from a
model of spinless resonant tunneling between Luttinger liquids \cite{KaneFisher,EggertAffleck1992},
allowing us to obtain the transport properties for arbitrary $r$ values from
previous knowledge accumulated for the Luttinger problem, as discussed in the
main text.


\section{IV. Quantum critical point as Majorana state}

In this final section, we show that the special value $r=1$, corresponding to a 
fine-tuned circuit impedance $R=h/e^2$ (close to our experimental value), allows
a detailed understanding of the nature of the quantum critical point. 
Technically, a standard refermionization~\cite{GiamarchiBook} of the tunneling 
term~(\ref{eq:LT}), leads to an effective interacting resonant Majorana level.

The mathematical equivalence to the Majorana model starts by performing a unitary transformation
\cite{emery92,komnik03}, $U=\exp[i(d^{\dagger}d-1/2) \phi_c(0) / \sqrt{2}]$, to
eliminate the $\phi_c$ field in the tunneling action, Eq. (\ref{eq:LT}), which
now reads:
\begin{eqnarray}
\nonumber
S_{\textrm{T}}&=&\int \!\!d\tau \Big[\frac{V_{S}}{\sqrt{2\pi a_0}}
e^{-i\frac{1}{\sqrt{2g}}\phi_{f}'(\tau)}d
+\frac{V_{D}}{\sqrt{2\pi a_0}}
e^{i\frac{1}{\sqrt{2g}}\phi_{f}'(\tau)}d \Big]\\
&& + {\rm c.c.}\;. 
\label{eq:ST}
\end{eqnarray}
This operation generates a new contact interaction between the dot and the 
phase field:
\begin{equation}
 H_{C}= -\pi v_F (d^{\dagger}d-1/2) \partial_x \phi_c(x=0) \;.
\end{equation}
For the special value $g=1/2$, we can exactly refermionize the problem by defining 
fictitious free fermion fields $\psi_{c}=e^{i\phi_{c}}/\sqrt{2\pi a_0}$ and
$\psi_{f}=e^{i\phi_{f}'}/\sqrt{2\pi a_0}$ (where we neglect the Klein
factors which do not play any role here). Electron waves in the leads and
phase fluctuations in the circuit are thus combined into these non-interacting
fermionic species. All the complexity of the tunneling process is now encoded in
the following expression:
\begin{eqnarray}
\label{eq:Hfinal}
H_\mathrm{Majorana}&\equiv&
H_T + H_\mathrm{Dot}+ H_C \\
\nonumber
&=& \big[ V_S \psi_f^\dagger(0) d +h.c.\big] + \big[V_D \psi_f(0) d +h.c.\big]\\
\nonumber
&& + \epsilon_d d^\dagger d  -\pi v_F \psi_c^\dagger(0)  \psi_c(0) 
(d^\dagger d - 1/2).
\end{eqnarray}
The peculiarity of this effective Hamiltonian is the presence of ``pairing''
terms, like $\psi_f(0) d$, in contrast to the initial tunneling term 
Eq.\,(\ref{eq:Htunnel}) where 
the number of the fermions is conserved. This structure motivates the 
introduction of a Majorana description of the local level, $d^\dagger=a-ib$
where $a$ and $b$ are real fermions satisfying $\{a,b\}=0$ and $a^2=b^2=1/4$.
The effective Hamiltonian~(\ref{eq:Hfinal}) is then readily expressed as
\begin{eqnarray}
\label{eq:HMajorana}
H_\mathrm{Majorana}&=&
(V_S-V_D)\big[ \psi_f^\dagger(0)-\psi_f(0)\big] a \\ 
\nonumber
&& + i (V_S+V_D) \big[ \psi_f^\dagger(0)+ \psi_f(0)\big] b \\
\nonumber
&& + 2 i \epsilon_d a b  - 2 i \pi v_F \psi_c^\dagger(0)  \psi_c(0) a b ;
\end{eqnarray}
we have, thus, arrived at Eq.~(1) of the main text. 

A special point of Hamiltonian~(\ref{eq:HMajorana}) can then be identified
when $V_S=V_D$ (symmetric tunneling amplitudes to source and drain) and 
$\epsilon_d=0$ (resonant condition), in which case the $a$ Majorana does
not hybridize to either the leads or the $b$ Majorana level; the latter
is on the other hand tunnel coupled to the fermion bath. In the absence
of the density interaction to the field $\psi_c$ (last term in
Eq.~(\ref{eq:HMajorana})), one ends up with a non-interacting Majorana
resonant level model, describing a critical state with fractional degeneracy
due to the perfect decoupling of the Majorana $a$ mode 
(the ground state entropy 
is then $S=\frac{1}{2}\log 2$). Gogolin and Komnik~\cite{komnik03} 
introduce an extra, fine-tuned Coulomb 
interaction between the quantum dot and the leads to get rid of
the Majorana interaction term. However, this simplifying hypothesis is not
appropriate in our case and leads to an incorrect description 
of the quantum critical point. Indeed, for non-interacting Majoranas,
the conductance at low temperature assumes a Fermi liquid behavior with
quadratic in $T$ approach to the unitary value, which corresponds to an exact
and unfortunate cancellation of the leading irrelevant operator discussed
in the main text. Having a finite Majorana interaction 
presents two interesting features: (i) the entropy of the critical state remains
unchanged at $S=\frac{1}{2}\log 2$ (the $a$ mode is not hybridized with the
continuum despite the density-density interaction~\cite{SenguptaGeorges,WongAffleck,ColemanIoffeTsvelik});
(ii) an anomalous non-Fermi liquid scattering rate is generated, with linear
behavior in temperature and voltage. This recovers the correct behavior due to the leading
irrelevant operator in the approach to the critical point~\cite{Zheng_new}. 

Note that the four-fermion interaction term in Eq.~(\ref{eq:HMajorana}) 
is too large to be quantitatively captured by perturbation theory, but rather drives 
the Majorana-level model into a strong coupling regime~\cite{ZitkoSimon} 
with universal scaling relations describing the quantum critical state in 
our system. Nevertheless, insight into the striking non-Fermi liquid behavior 
can be gained by calculating the conductance perturbatively in the interaction~\cite{Zheng_new}, as mentioned in the main text. 
The hybridized Majorana fermion $b$ has a
decreasing correlation function at long time, namely $G_b(t)=\big<b^+(0)b(t)\big>\propto 1/t$, 
identical to the behavior of the local (one-dimensional) bath Green's function
$G_{\psi_{c}(x=0)}(t)\propto 1/t$. However
the non-hybridized mode $a$ does not decay at long-times, 
$G_a(t)=\big<a^+(0)a(t)\big> \propto 1$.
The role of the interaction given by the last term in Eq.~(\ref{eq:HMajorana}) 
is to generate a self-energy for the hybridized mode $b$; at lowest order
in perturbation theory, it reads $\Sigma_b(t)\propto v_F^2 G_a(t) G_{\psi_c}^2(t)
\propto 1/t^2$. Now, the current-current correlation function of the original problem turns out to be proportional to the self-energy $\Sigma_b$ \cite{Zheng_new}. Its Fourier transform yields a dependence linear in frequency, and in this way a correction to the unitary conductance which is linear in temperature is produced, $1-G \propto T^1$. 
Away from the QCP, the Majorana mode $a$ becomes hybridized, and so
$\Sigma_b$ decays as $1/t^3$, which by Fourier transform leads to the expected 
quadratic-in-energy Fermi liquid behavior.

Finally, for $r$ close to but not exactly $1$, corresponding to our experimental
situation (the circuit impedance cannot be tuned \emph{in situ} but rather is
fixed by the lithographic process), one ends up with a Majorana model similar to
Eq.~(\ref{eq:HMajorana}) but with weakly interacting Luttinger 
leads~\cite{FabrizioGogolin,goldstein10,AffleckPrivatecom} with effective 
Luttinger parameter $\tilde{g} \propto r$.
This residual interaction within the continuum leads to a slight deviation from 
the linear approach to the unitary conductance, as we indeed observe in our data 
(see Fig.2a of the main text). Although the critical state is not precisely a 
Majorana bound state in that case, the associated ground state still possesses 
entropy 
$S=\frac{1}{2}\log(1+r)$  
associated with a non-trivial fractional degeneracy~\cite{AffleckPrivatecom,WongAffleck}.

\end{document}